# Wurtzite Effects on Spin Splitting of GaN/AlN Quantum Wells

Ikai Lo[*], W. T. Wang, M. H. Gau, S. F. Tsay, and J. C. Chiang[*]

Department of Physics, Center for Nanoscience and Nanotechnology, National Sun Yat-Sen University, Kaohsiung, Taiwan, Republic of China.

## Abstract

A new mechanism ($\Delta_{C1}$-$\Delta_{C3}$ coupling) is accounted for the spin splitting of wurtzite GaN, which is originated from the intrinsic wurtzite effects (band folding and structure inversion asymmetry). The band-folding effect generates two conduction bands ($\Delta_{C1}$ and $\Delta_{C3}$), in which *p*-wave probability has tremendous change when $k_z$ approaches anti-crossing zone. The spin-splitting energy induced by the $\Delta_{C1}$-$\Delta_{C3}$ coupling and wurtzite structure inversion asymmetry is much larger than that evaluated by traditional Rashba or Dresselhaus effects. When we apply the coupling to GaN/AlN quantum wells, we find that the spin-splitting energy is sensitively controllable by an electric field. Based on the mechanism, we proposed a *p*-wave-enhanced spin-polarized field effect transistor, made of $In_xGa_{1-x}N/In_yAl_{1-y}N$, for spintronics application.

I.  **Introduction**

Gate-controlled spin splitting in two dimensional electron system has been investigated in many *zinc-blende* III-V semiconductor quantum wells.[1,2] The gate-controlled spin splitting is arisen from the spin-orbit coupling[3] in zinc-blende structure with respect to inversion asymmetry. Carriers confined in asymmetric quantum wells will experience an effective magnetic field that may induce spin precession.[4] The manipulation of electron spins in a semiconductor is one of the key problems in the field of spintronics, in which additional degrees of freedom executed by electron spins are expected to play important roles in future nano-scaled electronic devices.[5,6] The spin splitting in *zinc-blende* III-V compound is induced either by a bulk inversion asymmetry of crystal potential (the $k^3$-term, called Dresselhaus effect),[7] or by a structure inversion asymmetry of electrostatic confinement potential (the linear-$k$ term, named Rashba effect).[8] Ganichev et al. have demonstrated the spin-orbital Hamiltonian of Rashba or Dresselhaus effects for *zinc-blende* InAs quantum well (QW) in terms of a $k$-dependent effective crystal magnetic field $\boldsymbol{B}_{\text{eff}}(\boldsymbol{k})$; e.g., $\boldsymbol{H}_{\text{SO}} = \boldsymbol{\sigma} \cdot \boldsymbol{B}_{\text{eff}}(\boldsymbol{k})$, where $\boldsymbol{k}$ is the electron wave vector and $\boldsymbol{\sigma}$ the vector of Pauli matrices. The presence of $\boldsymbol{B}_{\text{eff}}(\boldsymbol{k})$ implies that the spin orientation of electrons depends on the $k$-dependent Rashba and Dresselhaus terms.[4] Recently, Tsubaki et al.[9] and our group[10] independently observed a large spin-splitting energy (greater than 5 meV) in the 2DEG of GaN/AlGaN *wurtzite* heterostructures. The *wurtzite* GaN-based QW can be a potential candidate to realize the gate-controlled spin-polarized devices. However, the spin-splitting energy of *wurtzite* GaN, calculated by a traditional Rashba model (~1 meV), is much smaller than the measured values.[11,12] Reviewing the spin-orbital interaction in *wurtzite* semiconductors, Lew Yan Voon et al. have pointed out that, in addition to Dresselhaus $k^3$-term, there exists a linear-$k$ term spin-splitting energy caused by an intrinsic structure inversion asymmetry (SIA).[13] It was specifically pointed out that this linear-$k$ term arises from the weak $s$-$p_z$ mixing of conduction band at $\boldsymbol{k} = 0$ (i.e., $\Gamma_{C1}$ state) mainly due to the second Ga-N neighbor interactions and has been vigorously investigated since the 1950s.[14] However, Lew Yan Voon calculated the spin splitting due to the linear-$k$ and $k^3$ terms for wurtzite CdS and ZnO, and obtained the spin-splitting energy at small $k$-value less than 0.5 meV and 0.05 meV, respectively. It reveals that a new mechanism beyond the Rashba, Dresselhaus and second Ga-N neighbor interactions is needed to calculate the spin splitting of wurtzite GaN. For the wurtzite structure, the intrinsic SIA effect (i.e., the linear-$k$ term) exists even in an ideal wurtzite structure because of the long range interactions (mainly the second Ga-N neighbor interactions). Moreover, as pointed out by Lew Yan Voon et al., deviations from ideal structure (e.g., caused by crystal field or strain) can also lead to linear-$k$ terms in the spin-splitting energy.[13] Thus, for convenience, we define the net effect of the intrinsic SIAs mentioned above as the wurtzite structure inversion asymmetry (WSIA) effect. Another intrinsic wurtzite effect, band folding, has been reported and observed in III-Nitrides.[15] The band folding effect is created when the structure of III-Nitrides is transformed from cubic zinc-blende to hexagonal wurtzite. Two conduction bands ($\Delta_{C1}$ and $\Delta_{C3}$) are generated as the wave vector along Γ-L [111] direction of zinc-blende is folded back to Γ-A [001] direction of wurtzite. In this paper, we take these intrinsic wurtzite effects into account to calculate the spin-dependent full band structures of GaN and AlN by the method of linear combination of atomic orbitals (LCAO). We find that the band folding effect creates a $\Delta_{C1}$-$\Delta_{C3}$ coupling, and a large spin-splitting energy is induced in the conduction band by the $\Delta_{C1}$-$\Delta_{C3}$ coupling and WSIA effect. This spin-splitting energy is much larger than that evaluated by Rashba or Dresselhaus effects and very sensitive to an applied electric field. Based on the mechanism, we proposed a spin-polarized field effect transistor which can further integrate the wide band-gap materials to spintronic devices.

## II. Band Calculation

The spin-dependent full band structure is calculated, based on the method used by Kobayashi et al.,[16] but the Hamiltonian with spin-orbital interaction we used is a 32x32 matrix: $H_{32 \times 32} = H_0 + H_{SO}$. The first Brillouin zone and its high symmetric points are shown in the inset of Fig. 1. The nonzero matrix elements are the interactions between orbitals on the same atoms, and the nearest-neighboring atoms (including nearest Ga-Ga, N-N and Ga-N). We include the WSIA effect of GaN via the reduction of lattice parameter $c$, by changing the distance of ideal wurtzite $c$(ideal) = 5.136Å to wurtzite $c$(wz) = 5.111Å.[3,12] Here, the influence of the long range interactions (mainly the second-neighbor Ga-N interactions) is regarded as a perturbation term in our calculations and included in the nearest-neighbor parameters. Semi-empirical parameters are used to solve Schrodinger equation.[3,17-20] The full band structure of wurtzite GaN is shown in Fig. 1 (a). Here, $\Delta_{C1}$ and $\Delta_{C3}$ represent the lowest and second lowest conduction bands, respectively. For the two conduction bands, a band crossing occurs along Γ-A direction,[21] due to the fact that Γ-A direction is the highest symmetrical orientation ($k_x = k_y = 0$). The crossing point is a mathematical saddle point. The calculated density of states indicates that, near Γ-point, $\Delta_{C1}$ band is an $s$-like state, while $\Delta_{C3}$ band is a highly hybrid $sp^3$-state which has much higher $p$-wave probability. This can be understood because $\Gamma_{C3}$-point is indeed folded back from $L_{C1}$-point of zinc-blende. The full band structure of AlN and its density of states are also calculated. The calculated band-folding effects for *ideal* wurtzite GaN and AlN are shown in Fig. 1 (b) and (c), where $L_{ZB}$-point is the boundary of zinc-blende Brillouin zone along [111] direction and $A_{WZ}$-point is that of ideal wurtzite Brillouin zone along [001] direction. Because the direct lattice length of unit cell in zinc-blende along [111] is equal to one half of that in ideal wurtzite along [001], $A_{WZ}$-point is just at the mid-point of Γ-$L_{ZB}$ section. It is shown that $\Delta_{C1}$ and $\Delta_{C3}$ bands (the empty circles) are crossed each other in GaN, but the crossing does not occur in AlN. The crossing is the result of such a band structure, in which the energy of $L_{C1}$-point is higher than that of $\Gamma_{C1}$-point and, meanwhile, the maximum energy of $\Lambda_{C1}$ band (the solid line) of zinc-blende appears before $A_{WZ}$-point. This is the case of GaN. But as the maximum energy of $\Lambda_{C1}$ band appears after $A_{WZ}$-point, no crossing is observed (the case of AlN). $\Delta_{C1}$ and $\Delta_{C3}$ bands versus $k_z$ for different values of $k_x$ (at $k_y = 0$) are plotted in Fig. 2. It is noted that Δ-band is traditionally assigned to the band along Γ-A direction ($k_x = k_y = 0$) in wurtzite, but we assign it, more generally, to those bands parallel to the Γ-A direction; i.e., for $k_{//} = (k_y^2 + k_y^2)^{1/2}$ within zone boundary. As expected, an anti-crossing appears when the symmetry is broken by $k_{//} \neq 0$. As $k_x$ increases (i.e., along Γ-M direction), both $\Delta_{C1}$ and $\Delta_{C3}$ bands shift up. The opening gap between $\Delta_{C1}$ and $\Delta_{C3}$ bands at anti-crossing zone increases with increasing $k_x$, and reaches ~0.07 eV at $k_x = 0.5\ \pi/c$, where the spin-degeneracy for both $\Delta_{C1}$ and $\Delta_{C3}$ bands is obviously lifted off.

### III.  Origin of Spin Splitting and WSIA Effects

The spin-splitting energies of $\Delta_{C1}$ (the blue line) and $\Delta_{C3}$ bands (the green line) for wurtzite GaN, denoted as $\delta E_{WZ}(\Delta_{C1})$ and $\delta E_{WZ}(\Delta_{C3})$, are plotted against $k_z$ in Fig. 3(a) at $k_x = 0.12\pi/c$ and $k_y = 0$. $\delta E_{WZ}(\Delta_{C1})$ increases with increasing $k_z$ and then a tremendous jump (from 2.3 meV up to 5.7 meV) takes place at $k_z \sim 0.384\pi/c$ (marked anti-crossing). After the jump, it decreases with increasing $k_z$ to the zone boundary. On the contrary, $\delta E_{WZ}(\Delta_{C3})$ decreases with increasing $k_z$ followed by a tremendous drop (from 5.7 meV down to 2.3 meV) at $k_z \sim 0.384\pi/c$. It then increases with increasing $k_z$ to the zone boundary. The switch of $\delta E_{WZ}(\Delta_{C1})$ and $\delta E_{WZ}(\Delta_{C3})$ arises from the anti-crossing of $\Delta_{C1}$ and $\Delta_{C3}$ bands at $k_z = 0.384\pi/c$ due to the band folding effect. In Fig. 3(a), the arrows show the effect of the band folding. If there was no anti-crossing between the two conduction bands at $k_z = 0.384\pi/c$, we should observe two folding bands from $\Lambda_{C1}$ band of zinc-blende (it is noted that an anti-crossing should occur when the symmetry is broken at $k_x = 0.12\pi/c$). The first folding band goes from arrow (1) to arrow (2), and the second goes from arrow (3) to arrow (4). When the coupling between these two folding bands gets involved, an anti-crossing occurs. In this paper, we define $\Delta_{C1}$ band as the lowest conduction band (the solid lines), thus it goes from arrow (1) into the anti-crossing zone, and then goes to the dotted arrow (3'). Meanwhile, $\Delta_{C3}$ band, defined as the second lowest conduction band (the dotted lines), goes from the dotted arrow (4') into the anti-crossing zone, and then goes to arrow (2). Because the spin splitting is mainly contributed from the spin-orbital interaction of p-wave, we check the probability of the density of states for the hybrid $sp^3$-wave in $\Delta_{C1}$ and $\Delta_{C3}$ bands, and shown in Fig. 3(b). For $\Delta_{C1}$ (or $\Delta_{C3}$) band, the probability of $s$-wave is 96.2% (or 54.6%) at $k_z = 0$ and it decreases (or increases) as $k_z$ increases. An abrupt drop (or jump) occurs at $k_z = 0.384\pi/c$ and then increases (or decreases) to 60.6% for both bands at the zone boundary. Similarly, the probability of $p$-wave is 3.8% (or 45.4%) for $\Delta_{C1}$ (or $\Delta_{C3}$) band at $k_z = 0$ and increases (or decreases) with increasing $k_z$. An abrupt jump (or drop) occurs at $k_z = 0.384\pi/c$ and then decreases (or increases) to 39.4% for both bands at the zone boundary. Because the $p$-wave probability of $\Delta_{C1}$ band increases with $k_z$ before the anti-crossing, $\delta E_{WZ}(\Delta_{C1})$ therefore increases with $k_z$. It has an abrupt jump at the anti-crossing, and then decreases with increasing $k_z$, and so $\delta E_{WZ}(\Delta_{C1})$ decreases. On the contrary, the $p$-wave probability of $\Delta_{C3}$ band, and hence $\delta E_{WZ}(\Delta_{C3})$, decreases with increasing $k_z$ before the anti-crossing. It has an abrupt drop at the anti-crossing, and then increases with increasing $k_z$, and so $\delta E_{WZ}(\Delta_{C3})$ increases. Therefore the spin-orbital interaction of $p$-waves in $\Delta_{C1}$ and $\Delta_{C3}$ bands dominates the spin-splitting energies, $\delta E_{WZ}(\Delta_{C1})$ and $\delta E_{WZ}(\Delta_{C3})$. The details of the coupling bands within the anti-crossing zone are shown in Fig. 3(c). The spin orientation is rotated during the band mixing between $\Delta_{C1}$ and $\Delta_{C3}$ and forms the spin-mixing bands ($\Delta^{\pm}_{C1}$ and $\Delta^{\pm}_{C3}$). The large spin splitting is contributed from the $p$-wave probability of the spin-mixed conduction bands. It is noted that a large spin-splitting energy is also obtained in another spin-polarized conduction bands of InAs/GaSb zinc-blende QW, but its conduction band is coupled with the valence bands of GaSb barrier more than those of InAs well, due to the Type II band alignment (InAs conduction band edge lies below GaSb valence band edge).[22] The $\Delta_{C1}$-$\Delta_{C3}$ coupling is quite different from the coupling between conduction and valence bands in zinc-blende, which is traditionally taken into account for Rashba and Dresselhaus effects. We also calculate the spin-splitting energies for the two conduction bands of *ideal* wurtzite GaN, $\delta E_{ideal}(\Delta_{C1})$ and $\delta E_{ideal}(\Delta_{C3})$, shown in Fig. 3(a). It is shown that both $\delta E_{ideal}(\Delta_{C1})$ and $\delta E_{ideal}(\Delta_{C3})$ become smaller, as compared to $\delta E_{WZ}(\Delta_{C1})$ and $\delta E_{WZ}(\Delta_{C3})$, respectively. In our calculations, the WSIA effect is not included in the ideal wurtzite, but it is included in the real wurtzite by means of semi-empirical parameters. This indicates that the WSIA enhances the spin-splitting energy of wurtzite GaN.

### IV. Spin Splitting in QW and Device Application

The Hamiltonian for GaN/AlN QW with a long-range electric field is written as $H_{32\times32} = H_0 + H_{SO} + H_{PE}$, where $H_{PE} = V_{PE}(r)I$ is an operator for long-range electrical potential induced by either piezoelectric field or an applied external field, and $I$ is a 32x32 unitary matrix.[23] The electric fields in GaN well and AlN barrier are assumed linear-dependent on layer thickness and set to be 100mV/Å between two interfaces of the QWs. The calculation is carried out for the QWs with different well-thickness (from 1 to 10 layers). The QW ground subbands of $\Delta_{C1}$ and $\Delta_{C3}$ bands at Fermi wave vector, $k^F=(k_\parallel^F, k_z^F)$, are denoted as $\Delta_{C1}(k^F)$ and $\Delta_{C3}(k^F)$, respectively. The spin-splitting energy of $\Delta_{C1}(k^F)$ subband at Fermi wave vector of $k_\parallel^F = 0.12\pi/c$ is plotted against well-thickness for the cases of the ideal wurtzite QW (empty circles), the wurtzite QW (empty squares), the ideal wurtzite QW with electric field (solid circles), and the wurtzite QW with electric field (solid squares) in Fig. 4. The value $k_\parallel = (k_x^2 + k_y^2)^{1/2} = 0.12\pi/c$ is referred to the carrier concentration of 2DEG, $n_{2D} = 8.73\times10^{12}$ cm$^{-2}$. It is shown that the spin-splitting energy of $\Delta_{C1}(k^F)$ increases with decreasing well thickness. This is because the $p$-wave probability of $\Delta_{C1}(k^F)$ increases as the well thickness decreases (as shown in the inset of Fig. 4). The reduction of well thickness shifts up the ground subbands of $\Delta_{C1}(k^F)$ and $\Delta_{C3}(k^F)$, and hence the quantized wave vector $k_z^F$ moves toward the anti-crossing zone; e.g., see Fig. 3(a). When it approaches much closer to the anti-crossing zone, the $\Gamma_{C1}$-like $\Delta_{C1}(k^F)$ and $\Gamma_{C3}$-like $\Delta_{C3}(k^F)$ will strongly couple with each other, and then turn into two highly mixing states: a $p$-wave-enhanced $\Delta_{C1}(k^F)$ and a $p$-wave-reduced $\Delta_{C3}(k^F)$. This coupling explains why the spin-splitting energy of $\Delta_{C1}(k^F)$ is able to reach as large as 4.4 meV in Fig. 4. On the other hand, if the quantized wave vector $k_z^F$ goes into the anti-crossing zone, $\Delta_{C1}(k^F)$ will become more $p$-like than $\Delta_{C3}(k^F)$, and then a very large spin-splitting energy in $\Delta_{C1}(k^F)$ is expected. The mechanism of the spin splitting mentioned here is based on the band folding effect. Besides, by comparing the spin-splitting energies between wurtzite and ideal wurtzite structures in Fig. 4, it reveals that the WSIA effect can significantly enhance the spin-splitting energy of $\Delta_{C1}(k^F)$. The large spin splitting in GaN/AlN QW is caused by the switch of $\Delta_{C1}(k^F)$ from $\Gamma_{C1}$-like state (about 0.2% $p_z$ wave) to $\Gamma_{C3}$-state (about 45% $p_z$ wave): the former is mostly contributed by the weak $s$-$p_z$ mixing of long range interactions and the latter by the strong $s$-$p_z$ mixing of nearest-neighbor interactions. We therefore conclude that the band folding and WSIA effects dominantly contribute to the spin splitting of wurtzite GaN/AlN QWs. It is worthy to point out that the traditional Dresselhaus and Rashba effects (based on the 4-band model of hybrid $sp^3$ states) provide very satisfactory results for the spin splitting in zinc-blende materials, but when applied to the wurtzite structure the band folding and WSIA effects need to be taken into account. We call these intrinsic band folding and WSIA effects as *wurtzite effects*.

In the spin-polarized field effect transistor proposed by Datta and Das, the electron spins of two-dimensional electron gas, injected from source to drain ferromagnetic electrodes, are controllably rotated when passing through InGaAs/InAlAs channel due to the spin-orbital interaction.[2,4] If we replace the InGaAs/AlInAs QW with an $In_xGa_{1-x}N/In_yAl_{1-y}N$ QW for the 2DEG channel, a larger spin-splitting energy can be achieved due to the higher $p$-wave probability, described above. According to the diagram of band-gap energy versus lattice constant,[24] we can use wurtzite ZnO (lattice constant $a$ = 3.252 Å) as a substrate to grow an $In_xGa_{1-x}N/In_yAl_{1-y}N$ lattice-matched QW (e.g., x = 0.27, y = 0.36). From the bowing parameters of $In_xGa_{1-x}N$ and $In_yAl_{1-y}N$ alloys, the difference of band-gap energy in the lattice-matched QW can reach as large as $\Delta E_g$ = 1.2 eV. Because $L_{C1}$-point in zinc-blende is four-fold degenerate, while $\Gamma_{C3}$-point in wurtzite is non-degenerate. The device is, therefore, benefited by the band folding effect, which lifts off the four-fold degeneracy of L-point in zinc-blende and removes the complexity induced by the degeneracy. The physical properties of wide band-gap QWs are also superior to the narrow band-gap QWs for electronic device application. In addition to the large band-gap difference and greater spin-splitting energy, another advantage of the new spin-polarized field effect transistor is that its strain strength is adjustable by changing alloy composition. Because the strength of WSIA can be tuned by the strain, we can optimize the device performance by adjusting the indium composition (x) of the strain-layered QW for spin-polarized field effect transistor. Based on the new mechanism, we therefore proposed the *p*-wave-enhanced spin-polarized field effect transistor, made of $In_xGa_{1-x}N/In_yAl_{1-y}N$ quantum well, which can further integrate the wide band-gap materials to spintronic devices by means of nano-technology.[25,26]

## V.  Conclusion

We have developed a new mechanism based on the band folding and wurtzite structure inversion asymmetry effects for the spin splitting of GaN. The $\Delta_{C1}$-$\Delta_{C3}$ coupling is caused by the band folding effect, and a large spin-splitting energy is induced by this coupling and the wurtzite structure inversion asymmetry. The spin-splitting energy of GaN/AlN QW is sensitively controllable by an applied electric field. We proposed a *p*-wave-enhanced GaN-based spin-polarized field effect transistor, made of $In_xGa_{1-x}N/In_yAl_{1-y}N$, which can integrate the wide band-gap materials and spintronics for new nano-scaled devices.

This project is supported in parts by National Research Council of Taiwan and Core Facilities Laboratory in Kaohsiung-Pingtung area.

References

*for correspondence: ikailo@mail.phys.nsysu.edu.tw (Ikai Lo), Chiang@mail.phys.nsysu.edu.tw (J.C. Chiang).

Figure Captions:

Fig. 1. (Color) (a) The full band structure calculated by LCAO method. The band folding effect generates two conduction bands ($\Delta_{C1}$ and $\Delta_{C3}$), which are crossing at the saddle point. The inset shows the first Brillouin zone and its high symmetric points. (b) and (c) show the band-folding effect on *ideal* wurtzite GaN and AlN, respectively. After the band folding, the section of zinc-blende $\Lambda_{C1}$ band between $A_{WZ}$- and $L_{ZB}$-points is folded to be wurtzite $\Delta_{C3}$ band, and $L_{C1}$-point turns into $\Gamma_{C3}$-point.

Fig. 2. (Color) The plots of $\Delta_{C1}$ and $\Delta_{C3}$ bands in wurtzite GaN against $k_z$ for different values of $k_x$ at $k_y = 0$. The two bands cross each other at the saddle point ($k_x = k_y = 0$). Here, we use the same unit ($\pi/c$) for $k_x$- and $k_z$-axes ($a/c = 0.6153$), for convenience reason. As $k_x$ is not equal to 0, the asymmetry generates an anti-crossing between the two bands and creates an opening gap.

Fig. 3. (Color) (a) The spin-splitting energies for $\Delta_{C1}$ and $\Delta_{C3}$ bands of *wurtzite* and *ideal* wurtzite GaN are plotted against $k_z$ for $k_x = 0.12\pi/c$ and $k_y = 0$. The band-folding effect for $\Delta_{C1}$ and $\Delta_{C3}$ in *wurtzite* GaN is also described by the arrows. (b) The probabilities of density of states for *s*- and *p*-waves in the two conduction bands. (c) The details of the spin-polarized $\Delta^{\pm}_{C1}$ and $\Delta^{\pm}_{C3}$ bands in the vicinity of the anti-crossing zone.

Fig. 4. (Color) The spin-splitting energies of $\Delta_{C1}(\boldsymbol{k}^F)$ are plotted against the well-thickness from 1 to 10 layers at $k_{//}^F = 0.12\pi/c$, for the cases with and without electric field in ideal wurtzite and wurtzite GaN/AlN QWs. The probabilities of density of states for *s*- and *p*-waves in $\Delta_{C1}(\boldsymbol{k}^F)$ and $\Delta_{C3}(\boldsymbol{k}^F)$ are also shown in the inset.

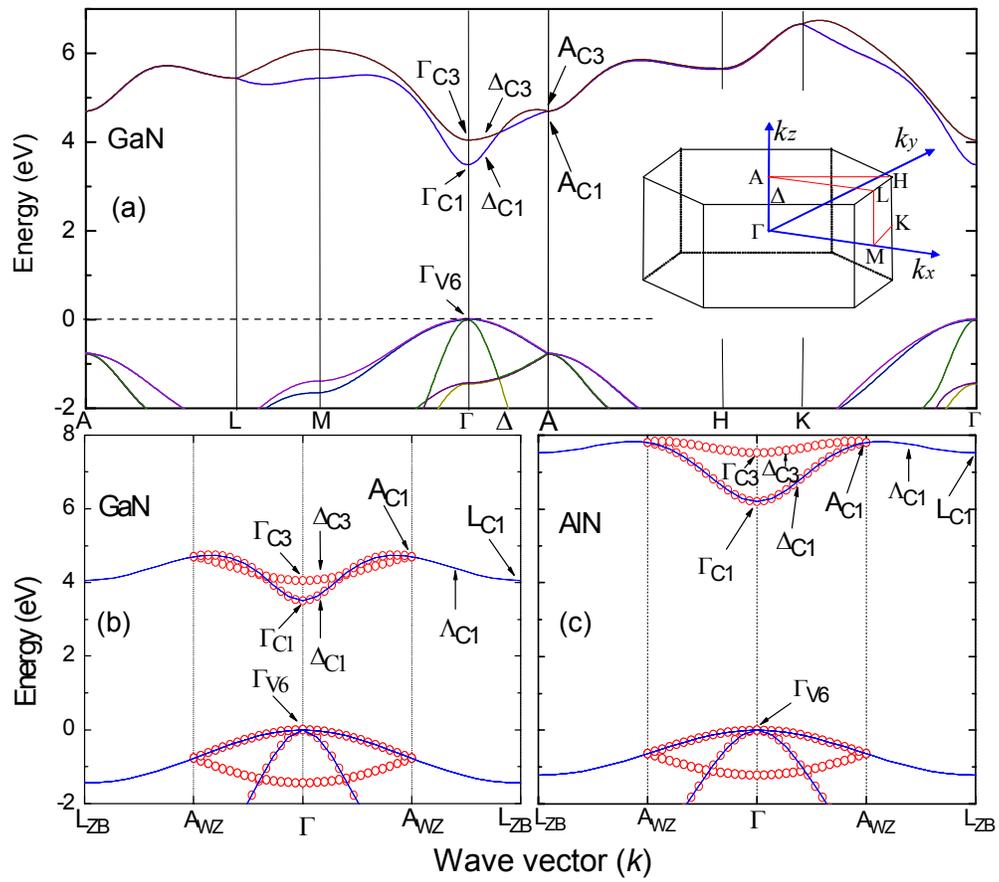

Fig. 1, Ikai Lo et al.

Fig. 2, Ikai Lo, et al.

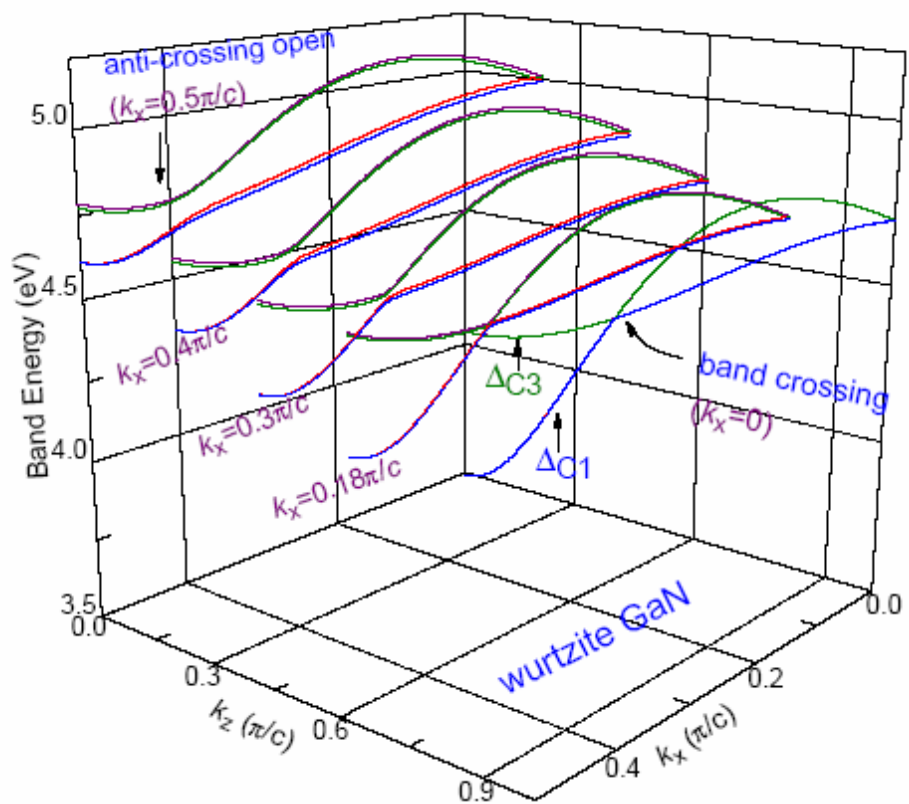

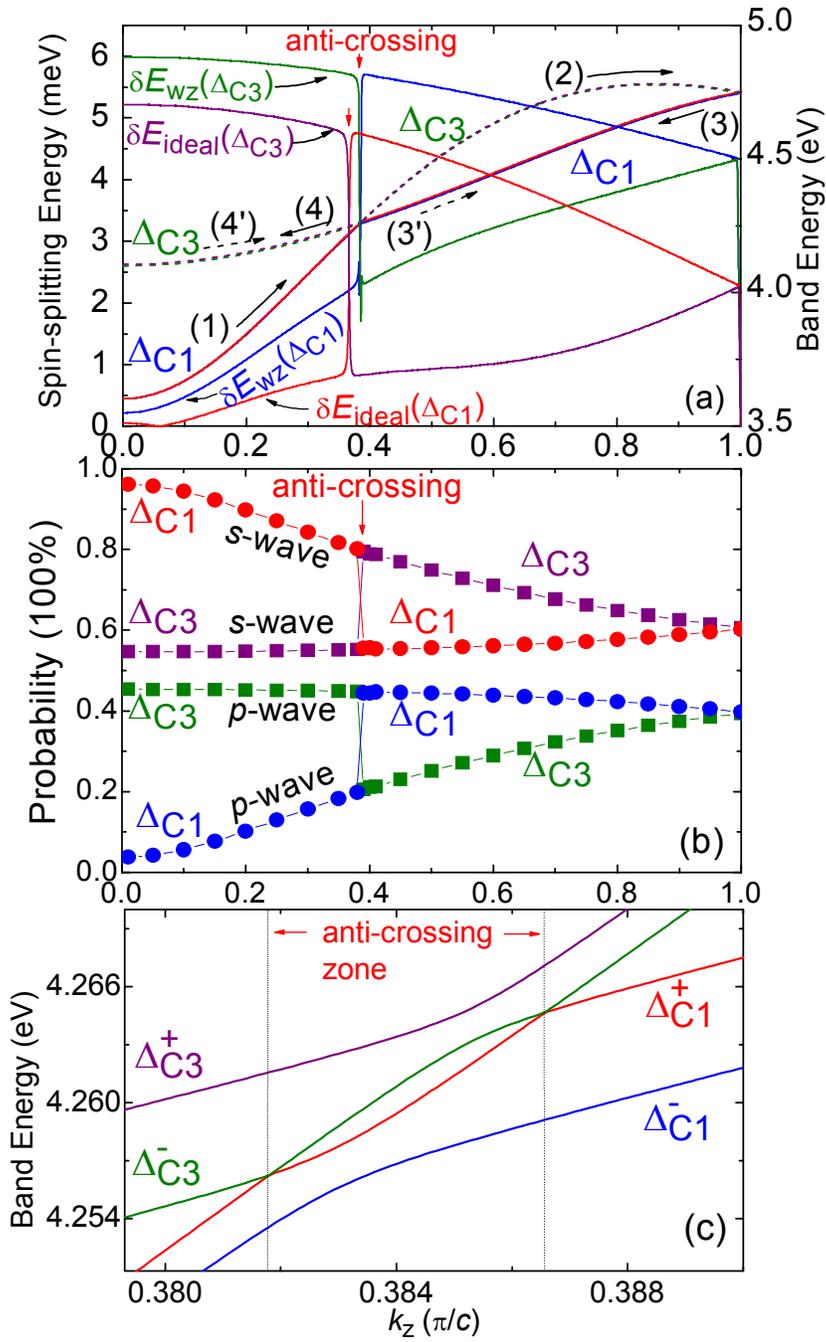

Fig. 3, Ikai Lo, et al.

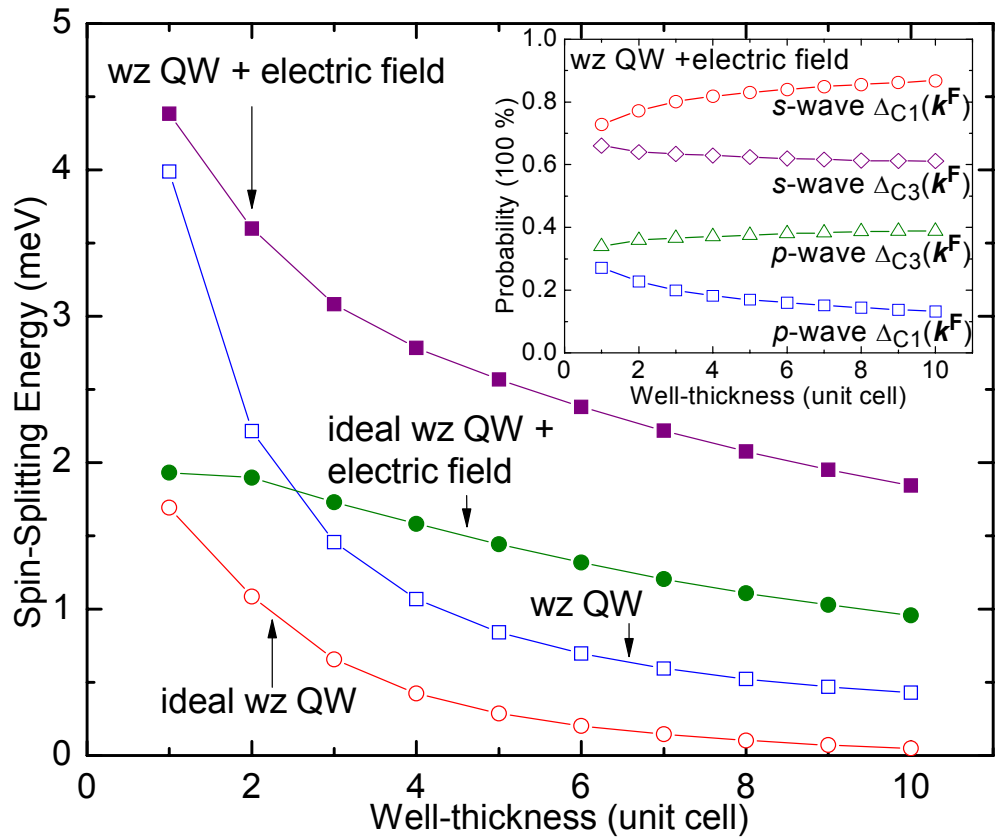

Fig.4, Ikai Lo et al.